\documentclass[twocolumn]{aastex62}

\newcommand{\kms}{km\,s$^{-1}$}
\newcommand{\cii}{[\ion{C}{2}]}
\newcommand{\ci}{[\ion{C}{1}]}
\newcommand{\oi}{[\ion{O}{1}]}
\newcommand{\mgii}{\ion{Mg}{2}}

\newcommand{\lsun}{$L_\sun$}
\newcommand{\msun}{$M_\sun$}

\newcommand{\msunyr}{$M_\sun$\,yr$^{-1}$}

\newcommand{\kkmspc}{K\,\kms\,pc$^2$}

\received{}
\revised{}
\accepted{}
\shorttitle{FIR Properties of Lensed Quasar J0439+1634}
\shortauthors{Yang et al.}
\begin{document}

\title{Far-Infrared Properties of the Bright, Gravitationally Lensed Quasar J0439+1634 at $z$=6.5}

\correspondingauthor{Jinyi Yang}
\email{jinyiyang@email.arizona.edu}

\author{Jinyi Yang}
\affil{Steward Observatory, University of Arizona, 933 N Cherry Ave, Tucson, AZ, USA}

\author{Bram Venemans}
\affil{Max-Planck Institute for Astronomy, K{\"o}nigstuhl 17, D-69117 Heidelberg, Germany}

\author{Feige Wang}
\affil{Department of Physics, University of California, Santa Barbara, CA 93106-9530, USA}

\author{Xiaohui Fan}
\affil{Steward Observatory, University of Arizona, 933 N Cherry Ave, Tucson, AZ, USA}

\author{Mladen Novak}
\affil{Max-Planck Institute for Astronomy, K{\"o}nigstuhl 17, D-69117 Heidelberg, Germany}

\author{Roberto Decarli}
\affil{Osservatorio Astronomico di Bologna, via Gobetti 93/3, I-40129 Bologna, Italy}

\author{Fabian Walter}
\affil{Max-Planck Institute for Astronomy, K{\"o}nigstuhl 17, D-69117 Heidelberg, Germany}

\author{Minghao Yue}
\affil{Steward Observatory, University of Arizona, 933 N Cherry Ave, Tucson, AZ, USA}

\author{Emmanuel Momjian}
\affil{National Radio Astronomy Observatory, P. O. Box O, Socorro, NM, 87801, USA}

\author{Charles R. Keeton}
\affil{Department of Physics and Astronomy, Rutgers University, Piscataway, NJ 08854}

\author{Ran Wang}
\affil{Kavli Institute for Astronomy and Astrophysics, Peking University, Beijing 100871, China}

\author{Ann Zabludoff}
\affil{Steward Observatory, University of Arizona, 933 N Cherry Ave, Tucson, AZ, USA}

\author{Xue-Bing Wu}
\affil{Kavli Institute for Astronomy and Astrophysics, Peking University, Beijing 100871, China}
\affil{Department of Astronomy, School of Physics, Peking University, Beijing 100871, China}

\author{Fuyan Bian}
\affil{European southern Observatory, Alonso de C\'ordova 3107, Casilla 19001, Vitacura, Santiago 19, Chile}



\begin{abstract}
We present IRAM/NOEMA, JCMT/SCUBA-2 and VLA observations of the most distant known gravitationally lensed quasar J0439+1634 at $z = 6.5$.
We detect strong dust emission, \cii\ 158 $\mu$m, \ci\ 369 $\mu$m, \oi\ 146 $\mu$m, CO(6--5), CO(7--6), CO(9--8), CO(10--9), H$_{\rm 2}$O $3_{\rm 1,2}-2_{\rm 2,1}$, and H$_{\rm 2}$O $3_{\rm 2,1}-3_{\rm 1,2}$ lines as well as a weak radio continuum. The strong \cii\ line yields a systemic redshift of the host galaxy to be $z=6.5188\pm0.0001$. The magnification makes J0439+1634 the far-infrared (FIR) brightest quasar at $z > 6$ known, with the brightest \cii\ line yet detected at this redshift. The FIR luminosity is (3.4$\pm$0.2)$\times$10$^{13}$ $\mu^{-1}$\,$L_{\odot}$, where $\mu$ $\sim$ 2.6 -- 6.6 is the magnification of the host galaxy, estimated based on the lensing configuration from {\em HST} imaging. We estimate the dust mass to be (2.2$\pm$0.1)$\times$10$^{9}$ $\mu^{-1}$\,\msun. The CO Spectral Line Energy Distribution using four CO lines is best fit by a two-component model of the molecular gas excitation. The estimates of molecular gas mass derived from CO lines and atomic carbon mass are consistent, in the range of 3.9 -- 8.9 $\times$10$^{10} \mu^{-1}$\,\msun. The [\ion{C}{2}]/[\ion{C}{1}], [\ion{C}{2}]/CO, and [\ion{O}{1}]/[\ion{C}{2}] line luminosity ratios suggest a photodissociation region model with more than one component. The ratio of H$_{\rm 2}$O $3_{\rm 2,1}-3_{\rm 1,2}$ line luminosity to $L_{\rm TIR}$ is consistent with values in local and high redshift ultra-/hyper-luminous infrared galaxies. The VLA observations reveal an unresolved radio continuum source, and indicate that J0439+1634 is a radio quiet quasar with R = 0.05 -- 0.17.
\end{abstract}
%



\section{Introduction} \label{sec:intro}
As the most luminous non-transient objects, quasars at $z >$ 6 are unique probes for the investigation of supermassive black holes (SMBHs) and their host galaxies at early cosmic time. Up to date, more than 150 quasars have been discovered at $z >$ 6, with the highest redshift at $z =$ 7.54 \citep{banados18}. Detections of such objects indicate the existence of up to (ten) billion $M_{\odot}$ SMBHs and place the strongest constraints on SMBH-galaxy coevolution at early epochs. A large fraction of $z>6$ quasars have been detected with significant [C\,{\sc ii}], CO and dust emission in the rest-frame far-infrared (FIR). The sub-millimeter/millimeter (submm/mm) observations suggest that the vast majority of the quasar host galaxies are intensely forming stars, at rates of a few 100 to 1000 $M_{\odot}$ yr$^{-1}$ \citep[e.g.][]{bertoldi03a,maiolino05,walter09,wang13,willott15,venemans16,venemans17a,venemans17b,decarli18,venemans18}, supported by large reservoirs of molecular gas, resulting in bright CO detections \citep[e.g.][]{bertoldi03b,walter03,walter04,wang11,venemans17c}.

Strong gravitational lensing provides a unique probe of the properties of quasars and their host galaxies. Lensing not only magnifies fluxes, but also stretches images, resulting in higher effective spatial resolution and better separation of the central quasar and host galaxy \citep{carilli03,peng06,treu10}. Lensed quasars at $z \sim$ 3 and 4 have been important tools for our understanding of interstellar medium (ISM) in high redshift quasar hosts \citep[e.g.,][]{weiss07}. At higher redshift ($z > 4.5$), lensed quasars could be powerful tools to study those distant supermassive black holes (SMBHs) and their host galaxies that are not easily resolved by current facilities.
However, at $z > 4.5$, only two lensed quasars at $z\sim 4.8$ \citep{mcgreer10,more11} were known until recently, although more than 200 quasars at $z = 4 - 6.4$ have been observed with {\em HST} at 0$\farcs$1 resolution \citep{richards04, richards06}. 
The lack of high-redshift lensed quasars has been attributed to either a reduced magnification bias due to a flat quasar luminosity function or a strong selection effect against lensed objects in morphology or color selection used in high-redshift quasar surveys.

Our new wide-area $z \sim 7$ quasar survey \citep{wang18, yang19} in a $\sim$ 20,000 $deg^{2}$ field has discovered the most distant known lensed quasar at $z = 6.51$, UHS J043947.08+163415.7 \citep[hereafter J0439+1634][hereafter Paper I]{fan19}. The \mgii-based redshift is $z = 6.511 \pm 0.003$. In Paper I, based on our {\em HST} FR782N and FR853N imaging which has a spatial resolution of $\sim$ 0$\farcs$075, we fit a singular isothermal ellipsoid lensing model and obtain the best-fit three-images lensing model of quasar.
The high magnification ($\mu_{\rm quasar} =$ 51.3$\pm$1.4 at optical wavelengths\footnote{This magnification factor is derived from {\em HST} imaging for the central quasar; the total magnification for 
the extended host galaxy is usually significantly smaller, see Section 3.}) 
makes this quasar the brightest observed quasar at $z > 6$ at optical and near-infrared wavelengths. Follow-up observations with 
the James Clerk Maxwell Telescope (JCMT) indicate that J0439+1634 is also the brightest far-infrared (FIR) emitter at $z > 6$ (Paper I). 
In this paper, we report new FIR and radio observations of J0439+1634 with the IRAM NOrthern Extended Millimeter Array (IRAM/NOEMA), JCMT, and Karl G. Jansky Very Large Array (VLA) and the detections of gas, dust emission, and radio continuum emission. 
We describe the observations in Section 2. The high quality detections of dust continuum and various molecular and atomic emission lines allow detailed studies of the inter stellar medium in the quasar host, which are presented in Section 3. The summary is given in Section 4. 
All results below refer to a $\Lambda$CDM cosmology with parameters $\Omega_{\Lambda}$ = 0.7, $\Omega_{m}$ = 0.3, and H$_{0}$ = 70 $\rm km ~s^{-1} ~Mpc^{-1}$. 

\section{Observations}

\subsection{NOEMA Observations}
We used IRAM/NOEMA to detect the [C\,{\sc ii}], CO(10--9), CO(9--8), CO(7--6), CO(6--5), [C\,{\sc i}] 369 $\mu$m (hereafter [C\,{\sc i}]), \oi\ 146 $\mu$m (hereafter \oi), and water emission lines from quasar host through two programs in 2018 Summer and Winter. The observations were carried out from 2018 July to December with the array in D and C configurations, using 7--10 antennas. We used the PloyFiX correlator which has two 7.744 GHz sidebands separated by 7.744 GHz (centers are separated by 15.488 GHz). The NOEMA data have been reduced using the latest version of the GILDAS\footnote{http://www.iram.fr/IRAMFR/GILDAS} software.

For the first program in 2018 Summer, we designed three set-ups. The CO(6--5), CO(7--6), and [C\,{\sc i}] lines were covered by a single setting with receiver 1 (3 mm) tuned at 107.410 GHz. The CO(9--8), CO(10--9), H$_{\rm 2}$O $3_{\rm 1,2}-2_{\rm 2,1}$, and H$_2$O $3_{\rm 2,1}-3_{\rm 1,2}$ lines were observed with receiver 2 (2 mm) tuned at 153.395 GHz. The [C\,{\sc ii}] line was observed in receiver 3 (1 mm) with tuning frequency at 253.068 GHz.
Quasars 0507+179, J0440+146 and 0446+112 were used as amplitude and phase calibrators. Radio sources 3C84 and 3C454.3 were observed as bandpass calibrators. Stars MWC349 and LKHA101 were used to set the absolute flux density scale.
The total on-source integration time in the receivers 1, 2, and 3 were 1.6, 2.8, and 1.3 hours (8 antenna equivalent), respectively. 
The synthesized beams are $4\farcs9 \times 1\farcs9$, $3\farcs1 \times 2\farcs4$, and $6\farcs2 \times 3\farcs2$ in the 1 mm, 2 mm, and 3 mm bands, respectively. 
The final 1 mm, 2 mm, and 3 mm cubes reach the sensitivity of 2.2 mJy beam$^{-1}$, 0.7 mJy beam$^{-1}$, and 0.5 mJy beam$^{-1}$ per 50 km s$^{-1}$ channel (1-$\sigma$).
In the second program in 2018 Winter, the \cii\ and \oi\ lines were covered by one setting with the receiver 3 (1mm) with a tuning frequency of 270.244 GHz. The on-source integration time in this observation is 3.13 hours (9 antenna equivalent). The synthesized beam is $1\farcs2 \times 0\farcs7$ and final data cube reaches a 1.5 mJy beam$^{-1}$ per 50 km s$^{-1}$ sensitivity (1-$\sigma$).

From the NOEMA observations, we obtained high quality detections of the CO(6--5), CO(7--6), CO(9--8), CO(10--9), \cii, \ci, \oi, H$_2$O $3_{\rm 1,2}-2_{\rm 2,1}$, and H$_2$O $3_{\rm 2,1}-3_{\rm 1,2}$ emission lines (see Figure 1 \& 2). The dust continuum is significantly detected in all cubes. All the NOEMA detections are spatially unresolved.
Measured line fluxes and continuum flux densities are summarized in Table 1 and Section 3. We only include statistical errors and the systematic flux density calibration uncertainties (of the order of $\sim$15\%) are not taken into account.

\subsection{JCMT Observations}
We observed the 450 $\mu$m (observed frame, 666 GHz) and 850 $\mu$m (353 GHz) dust continuum from J0439+1634 using the Submillimetre Common-User Bolometer Array 2 \citep[SCUBA-2;][]{holland13} on the JCMT in 2018 February. The observations were carried out in Band 2 weather conditions (i.e., 0.05 $< \tau_{225 GHz} <$ 0.008), and in $''$CV DAISY$''$ mode. The effective beam size of SCUBA-2 has a diameter of 9$\farcs$8 at 450 $\mu$m and 14$\farcs$6 at 850 $\mu$m. Our target was observed for 1.1 hours on source time (two 34 min scans). The data were reduced using the STARLINK SCUBA-2 pipeline for faint point sources \citep{chapin13}. The 1$\sigma$ sensitivities in 450 $\mu$m and 850 $\mu$m are 29.12 mJy beam$^{-1}$ and and 1.69 mJy beam$^{-1}$, respectively. 
From the JCMT observations, we obtained an 850 $\mu$m flux density of 26.20$\pm$1.68 mJy and a 2$\sigma$ signal of 65.71$\pm$29.12 mJy at 450 $\mu$m, which is measured at the peak flux pixel of the 850 $\mu$m detection.

\subsection{VLA Observations}
We observed the radio continuum emission from J0439+1634 using VLA DDT time in S band (2--4 GHz) with the A configuration. 
The observed reference frequency of 3 GHz corresponds to the rest-frame frequency of 22.6 GHz for this quasar.
The 8-bit samplers were utilized to cover the 2 GHz span of the S-band receiver. The total telescope time was 3 hours (2.1 hours on-source).
We used J0440+1437 for complex gain calibration and 3C138 for bandpass and flux density scale calibration.
The data were reduced by the the Common Astronomy Software Applications package (CASA) and the VLA calibration pipeline.
The synthesized beam was $0\farcs79 \times 0\farcs69$, and the 1$\sigma$ rms noise of continuum image was 4.5 $\mu$Jy beam$^{-1}$.
The VLA 3 GHz image shows an unresolved radio continuum source with a flux density of 28.9$\pm$4.5 $\mu$Jy.

\begin{figure*}
\centering 
\epsscale{1.1}
\plotone{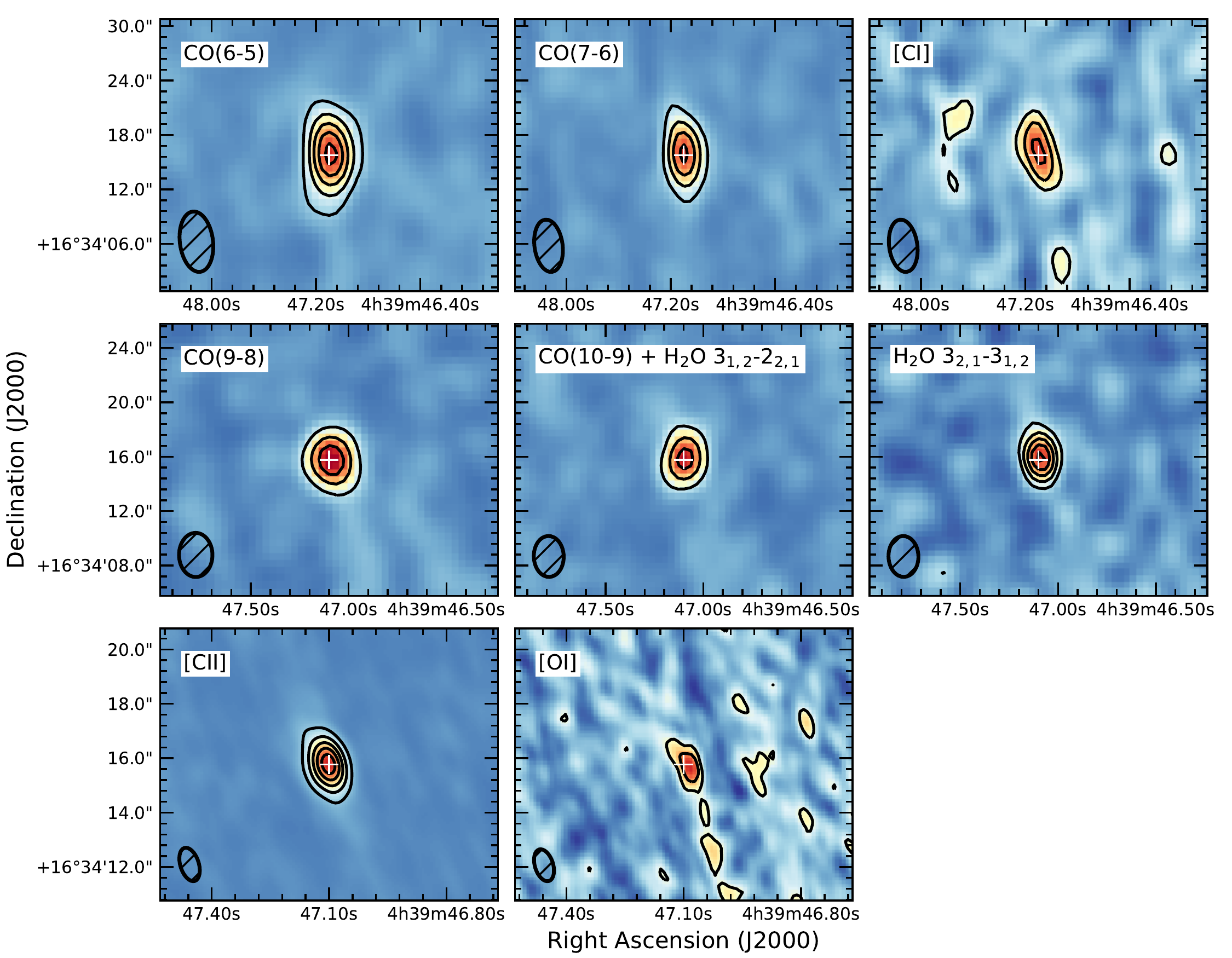} 
\plotone{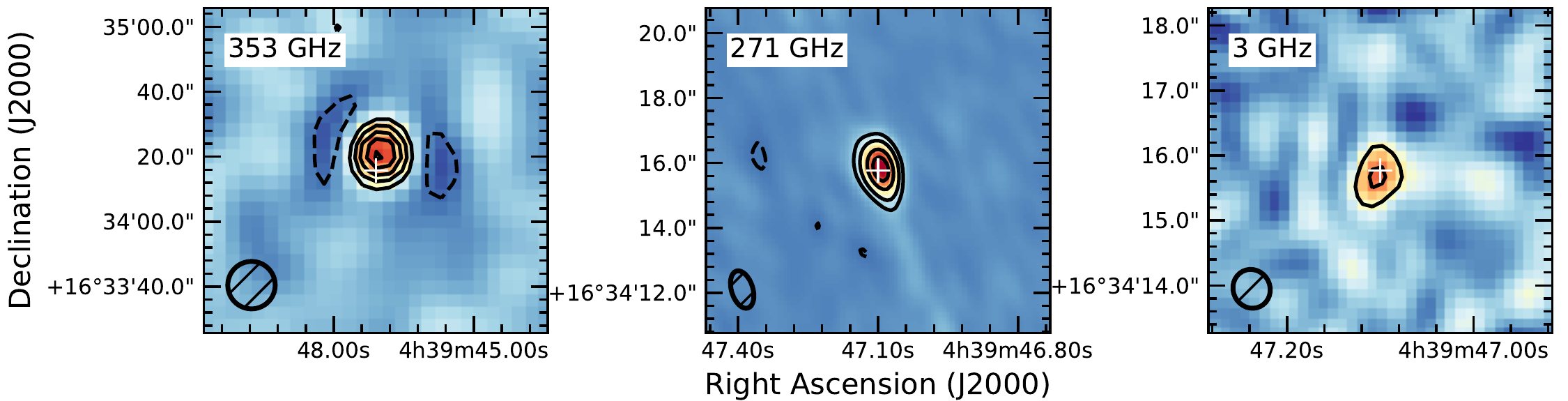} 
\caption{The maps of lines, FIR continuum, and radio continuum emissions. The line intensity map is obtained by integrating the continuum-subtracted data cubs over the velocity of each line. The white cross in each map denotes the position fitted from [C\,{\sc ii}], which is consistent with the optical coordinates.
The black contours in CO(6--5), CO(7--6), CO(9--8), and CO(10--9) maps are [3, 6, 9, 12, 15, 18] $\times \sigma$. The contours in [C\,{\sc i}], [O\,{\sc i}], and H$_2$O $3_{\rm 2,1}-3_{\rm 1,2}$ maps are [2, 3, 4, 5, 6] $\times \sigma$. The counters in the map of [C\,{\sc ii}] are [5,10,15, 20, 25, 30, 35] $\times \sigma$. The values of $\sigma$ in these line maps are, from upper-left (CO(6--5)) to bottom-right ([O\,{\sc i}]), 0.1, 0.1, 0.1, 0.2, 0.2, 0.2, 0.4, and 0.3 mJy \kms. 
In the maps of continuum emission, the contours are [3, 6, 9, 12, 15] $\times \sigma$ for S$_{\rm 353 GHz}$ ($\sigma$ = 1.7 mJy) and S$_{\rm 3 GHz}$ ($\sigma$ = 0.004 mJy ) and [10, 20, 40, 60, 80] $\times \sigma$ for S$_{\rm 271 GHz}$ ($\sigma$ = 0.2 mJy). The black dashed lines are --3 $\sigma$.
Here we only plot the S$_{\rm 271 GHz}$ as a representative of dust continuum emissions detected from NOEMA observations since all detections are spatially unresolved and the S$_{\rm 271 GHz}$ map has the smallest beam size. The peak of S$_{\rm 353 GHz}$ emission locates $\sim 3''$ away from the optical coordinates. Considering that the beam size of S$_{\rm 353 GHz}$ map is 14.6$''$ and there is no nearby source detected in all other FIR continuum maps, we believe that the S$_{\rm 353 GHz}$ emission is all from J0439+1634.}
\label{fig:spectrum}
\end{figure*}

\begin{figure*}
\centering 
\epsscale{1.0}
\plotone{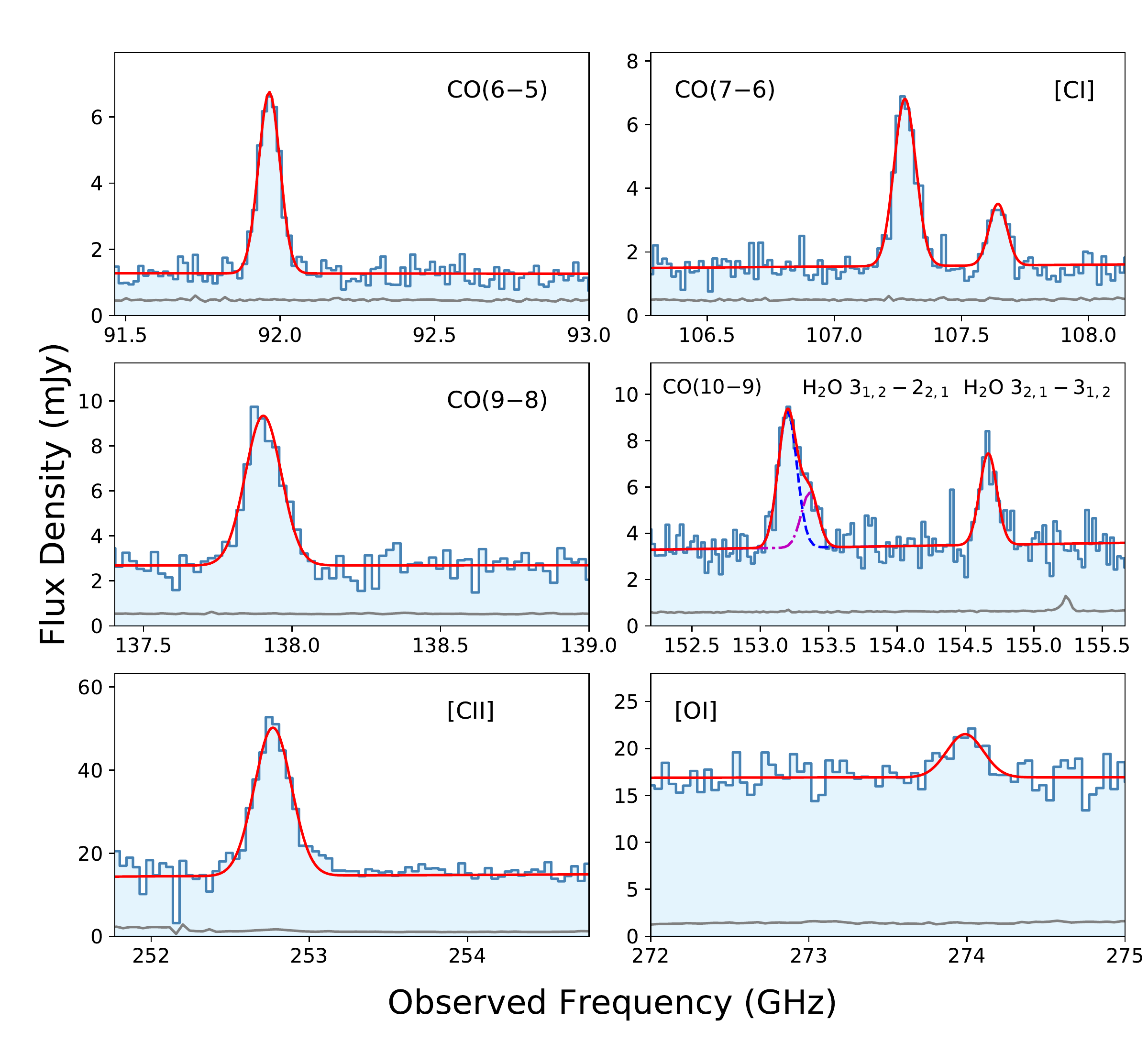} 
\caption{NOEMA spectra of the CO(6--5), CO(7--6), CO(9--8), CO(10--9), [C\,{\sc i}], [C\,{\sc ii}], [O\,{\sc i}], H$_2$O $3_{\rm 1,2}-2_{\rm 2,1}$, H$_2$O $3_{\rm 2,1}-3_{\rm 1,2}$ emission lines, and the underlying continuum of J0439+1634, extracted from the peak pixel in each data cube. The bin size is $\sim$50\,\kms. The grey lines show the uncertainties of the spectra. The [C\,{\sc ii}] line is the combination of two NOEMA observations. The red solid lines are the fits to spectra including a Gaussian profile for the emission lines plus continuum. The CO(10--9) line is blended with H$_2$O 3(1,2)--2(2,1). A two-Gaussian fit is used for the blended line. All measured values are reported in Table 1. These are the brightest lines observed from a $z > 6$ quasar host.}
\label{fig:spectrum}
\end{figure*}

\begin{table}%[!t]
%\begin{center}
\caption{Observed and Derived Properties of J0439+1634}%T-1
\footnotesize
%\centering
\begin{tabular}{lc}
\hline\hline
R.A.\ (J2000) & 04$^\mathrm{h}$39$^\mathrm{m}$47$.\!\!^\mathrm{s}$10 \\
Decl. (J2000) & $+$16$^\circ$34$^\prime$15\farcs77 \\
$z_\mathrm{[CII]}$   &   6.5188$\pm$0.0002 \\
$F_\mathrm{[CII]}$ (Jy\,\kms)   &    11.7$\pm$0.8\\
FWHM$_\mathrm{[CII]}$ (\kms) & 328.1$\pm$18.0\\
$F_\mathrm{CO(6-5)}$ (Jy\,\kms)  & 1.5$\pm$0.1\\
FWHM$_\mathrm{CO(6-5)}$ (\kms)  & 266.0$\pm$15.8\\
$F_\mathrm{CO(7-6)}$ (Jy\,\kms)  &1.5$\pm$0.1 \\
FWHM$_\mathrm{CO(7-6)}$ (\kms)   & 282.7$\pm$18.0\\
$F_\mathrm{CO(9-8)}$ (Jy\,\kms)   & 2.1$\pm$0.2\\
FWHM$_\mathrm{CO(9-8)}$ (\kms)   & 312.5$\pm$22.4\\
$F_\mathrm{CO(10-9)}$ (Jy\,\kms)   & 1.9$\pm$0.2\\
FWHM$_\mathrm{CO(10-9)}$\tablenotemark{a} (\kms)   &312.5$\pm$17.9 \\
$F_\mathrm{[CI]}$ (Jy\,\kms)  & 0.5$\pm$0.1\\
FWHM$_\mathrm{[CI]}$ (\kms)  & 232.0$\pm$36.6\\
$F_\mathrm{[OI]}$ (Jy\,\kms)  & 1.4$\pm$0.3\\
FWHM$_\mathrm{[OI]}$ (\kms)  & 298.9$\pm$56.1\\
$F_{\mathrm{H}_2\mathrm{O(312-221)}}$ (Jy\,\kms) &0.9$\pm$0.2 \\
FWHM$_{\mathrm{H}_2\mathrm{O}(312-221)}$\tablenotemark{b} (\kms)  & 288.4$\pm$42.5\\
$F_{\mathrm{H}_2\mathrm{O}(321-312)}$ (Jy\,\kms)  & 1.1$\pm$0.2\\
FWHM$_{\mathrm{H}_2\mathrm{O}(321-312)}$ (\kms)  & 288.4$\pm$44.4\\
S$_{666\ \rm GHz}$ (mJy)  &  $<$ 87.3\\
S$_{353\ \rm GHz}$ (mJy)  &  26.2$\pm$1.7 \\
 S$_{271\ \rm GHz}$  (mJy)  &  16.9$\pm$0.1 \\
 S$_{255\ \rm GHz}$  (mJy)  &  15.5$\pm$0.1 \\
 S$_{239\ \rm GHz}$  (mJy)  &  14.0$\pm$0.1 \\
 S$_{155\ \rm GHz}$  (mJy)  &  3.5$\pm$0.04 \\
 S$_{139\ \rm GHz}$  (mJy)  &   2.7$\pm$0.03 \\ 
 S$_{109\ \rm GHz}$  (mJy)  &   1.6$\pm$0.03 \\
 S$_{93\ \rm GHz}$  (mJy)  &   1.3$\pm$0.03 \\
 S$_{3\,\mathrm{GHz}}$ ($\mu$Jy) & 28.9$\pm$4.5 \\
\hline
$\mu$\tablenotemark{c} & 2.6--6.6\\
$L_\mathrm{FIR}$ ($L_{\odot}$)& (3.4$\pm$0.2)$\times$10$^{13}$ $\mu^{-1}$\\
$L_\mathrm{TIR}$ ($L_{\odot}$) & (4.8$\pm$0.2)$\times$10$^{13}$ $\mu^{-1}$\\
$L_\mathrm{[CII]}$ ($L_{\odot}$) & (1.2$\pm$0.1)$\times$10$^{10}$ $\mu^{-1}$\\
$L^\prime_\mathrm{[CII]}$ (\kkmspc) & (5.7$\pm$0.4)$\times$10$^{10}$ $\mu^{-1}$\\
$L_\mathrm{[CI]}$ ($L_{\odot}$) & (2.0$\pm$0.4)$\times$10$^{8} $ $\mu^{-1}$\\
$L^\prime_\mathrm{[CI]}$ (\kkmspc) & (1.2$\pm$0.2)$\times$10$^{10}$ $\mu^{-1}$\\
SFR$_\mathrm{TIR}$ (\msunyr) & 7080 $\mu^{-1}$\\
SFR$_\mathrm{[CII]}$ (\msunyr) & 1000 -- 6000 $\mu^{-1.18}$\\
$M_d$ (\msun) & (2.2$\pm$0.1)$\times$10$^{9}$ $\mu^{-1}$\\
$M_{\rm CI}$ (\msun) & (2.6$\pm$0.5)$\times$10$^{7}$ $\mu^{-1}$\\
$M_{\rm C^+}$ (\msun) & (3.7$\pm$0.3) $\times 10 ^{7} \mu^{-1}$\\
$M_{\rm H_2, CO}$ (\msun)\tablenotemark{d}& 5.4 $\times$10$^{10}$ $\mu^{-1}$\\ 
$M_{\rm H_2, CI}$ (\msun )\tablenotemark{e}&  (3.9 -- 8.9) $\times$10$^{10}$ $\mu^{-1}$ \\
\hline
\end{tabular}
\tablenotetext{a}{\scriptsize We fix the FWHM of CO(10--9) line to that measured from CO(9--8) line.}
\tablenotetext{b}{\scriptsize We fix the FWHM of H$_{\rm 2}$O $3_{\rm 1,2}-2_{\rm 2,1}$ line to that measured from H$_{\rm 2}$O $3_{\rm 2,1}-3_{\rm 1,2}$ line.}
\tablenotetext{c}{\scriptsize $\mu$ is the lens magnification of host galaxy (see Section 3).}
\tablenotetext{d}{\scriptsize Molecular gas mass derived from the CO(1--0) luminosity, assuming a luminosity-to-gas mass conversion factor of $\alpha$ =  0.8$M_{\odot}$(\kkmspc)$^{-1}$.}
\tablenotetext{e}{\scriptsize Molecular gas mass derived from the atomic carbon mass, assuming a carbon abundance of X[C I]= (8.4$\pm$3.5) $\times 10^{-5}$.}
%\end{center}
\end{table}

\section{The Host Galaxy of J0439+1634}
The NOEMA spectra are shown in Figure 2 and all detections are summarized in Table 1. The detections of multi-emission lines and dust continuum at multiple frequencies allow us to investigate the properties of gas and dust in the host galaxy of J0439+1634. All measurements of emission lines and derived properties such as gas mass, dust mass, and star formation rate (SFR) are also listed in Table 1. The position of the quasar host derived from the \cii\ emission, R.A. = 04$^\mathrm{h}$39$^\mathrm{m}$47$.\!\!^\mathrm{s}$10 and Decl. = $+$16$^\circ$34$^\prime$15\farcs77, is consistent with the optical location of the quasar (Paper I).

J0439+1634 is a compact lensed quasar with $\sim$ 0$\farcs$2 separation, and has only been resolved in the {\em HST} image with a spatial resolution of 0$\farcs$075 (Paper I). 
The {\em HST} image was designed to image quasar's Ly$\beta$ emission line.
The two-orbit narrow-band imaging is relatively shallow and is dominated by emission from the central AGN. Therefore, 
 we can only model the lensing configuration of quasar base on {\em HST} image (see details in Paper I). The lensing configuration of host galaxy, in particular, the total magnification could be significantly different since the quasar is a point source, while the host galaxy is an extended source with unclear morphology, inclination angle, and position angle.
The source was not spatially resolved in our NOEMA, JCMT or VLA observations. Therefore, all measurements of intrinsic properties of the host galaxy are made based on the simulated lensing model of host galaxy and the simulation is based on the lensing configuration of quasar. 

We use the isothermal ellipsoid lens model from \cite{fan19}, including uncertainties sampled with Markov chain Monte Carlo methods. We assume the host galaxy as an exponential disk centered on the inferred position of the quasar. We set a grid of effective radius (0.1$''$ to 0.25$''$) based on the typical size of quasar host galaxies at this redshift \citep{wang13} and also assume a range of the projected ellipticity (0.0--0.8) with random position angles. 
The median magnification of the host galaxy ranges from 2.6 to 4.9 across the grid of effective radius and ellipticity; over the whole grid, the 95\% confidence interval for the magnification is 2.6--6.6.
Note that the total magnification of the host galaxy is significantly smaller than that of the central quasar ($\mu_{\rm quasar} \sim 51$) because the host galaxy is extended. 
For the rest of the paper, we will use $\mu$ as the magnification factor of the host galaxy. 
Our forthcoming high resolution ALMA imaging will directly resolve the host galaxy into a lensed arc, and allow detailed lensing modelling to measure the size and shape of the quasar host galaxy, and to derive its magnification factor accurately.

In a lensed system, the different angular size of extended emissions from different lines or continuum 
(e.g., if the dust emission is more compact than the \cii\ emitting region) 
could result in differential lensing effect. 
Differential lensing could also distort the velocity profile of emission lines, depending on how the kinematic structure of the source couples with the lensing magnification pattern \citep{rivera18,yang19}. The emission lines of J0439 from NOEMA observations are well fitted by a single Gaussian profile except for the blended CO(10--9) and H$_2$O $3_{\rm 1,2}-2_{\rm 2,1}$ and the possible wings of \cii\ line (see Figure 2 and Section 3.1). Therefore, we do not include any discussion of this effect in this paper, and until we resolve the host we assume the same magnification of all FIR emission. The intrinsic properties measured in this paper, especially the ratios between different parameters, could nevertheless be affected by differential lensing. 
Giving these complications, in this paper, we simply report the {\em observed} values of quasar host galaxy properties with a common magnification factor $\mu$.

\subsection{Continuum Emission and Infrared (IR) Luminosity}

\begin{figure}
\centering
\epsscale{1.3}
\plotone{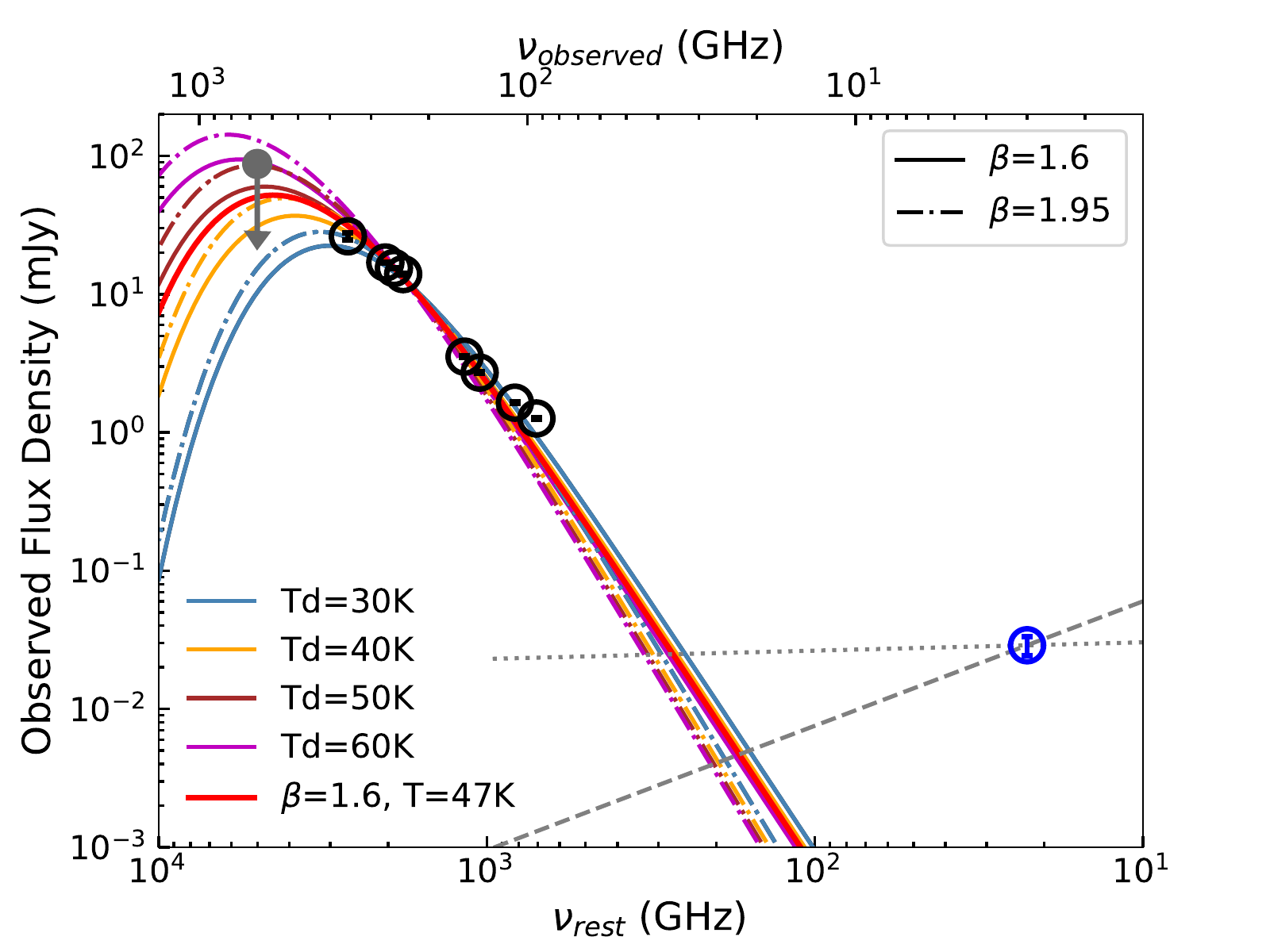} 
\caption{FIR and radio spectral energy distribution of J0439+1634. Black circles represent measurements from the JCMT and NOEMA observations of the dust continuum. The grey point with arrow denote the 3$-\sigma$ upper limit at 666 GHz. The blue circle is the 3 GHz detection from the VLA, which shows the radio emission. Two grey lines are the radio continua with different slopes, $\alpha = -0.06$ (dotted) and $\alpha = -0.9$ (dashed), assuming the slope range based on previous results of quasar J0100+2802 \citep{wang16}.
The red line is the typical dust emission model with $T_{d}$ = 47 K and $\beta$ = 1.6 \citep{beelen06}. The dot-dashed lines are models with $\beta =1.95$. A series of models with different $T_{d}$ are plotted for comparison. With fixed $\beta$ = 1.6, the best-fitting $T_{d}$ is 39 $\pm$9 K, which changes to $T_{d}$ = 29 $\pm$ 6 K with  $\beta = 1.95$. } 
\label{fig:dust}
\end{figure}

We detect dust continuum at observed frame frequencies of 352.7 GHz from JCMT, and 271 GHz, 254 GHz, 239 GHz, 155 GHz, 139 GHz, 109 GHz, and 93 GHz from NOEMA. The dust continuum from the NOEMA observations used in this section are measured by averaging spectra over all channels within the line free regions in each sideband. The dust continuum is detected with high signal-to-ratio in the NOEMA spectra, as shown in Figure 1 and Figure 2. All flux densities are listed in Table 1. The JCMT observation at 666.2 GHz show a 3$-\sigma$ upper limit.

To estimate the far-infrared (FIR, rest-frame 42.5--122.5 $\mu$m) luminosity $L_{\rm FIR}$, total infrared (TIR, rest-frame 8--1000 $\mu$m) luminosity $L_{\rm TIR}$ and the dust mass, we fit the dust emission with an optically thin grey body following the literature \citep[e.g.,][]{beelen06}. We fit a set of models with different dust temperatures $T_{d}$ and fixed the emissivity index $\beta$ = 1.6 or 1.95, as shown in Figure 3. We take the effect of the cosmic microwave background (CMB) on the dust emission into account \citep[e.g.,][]{dacunha13}. We exclude the 666.2 GHz upper limit for the model fitting.
We obtain a best-fit (least $\chi^{2}$) dust temperature of $T_{d}$  = 39 $\pm$9 K at fixed $\beta$=1.6, consistent with the canonical value of $T_{d}$ = 47 K \citep{beelen06}. If we fix the $T_{d}$ to 47 K, the best-fitting value of $\beta$ is 1.4$\pm$0.2.
The temperature strongly depends on the flux densities measured near the peak of the dust spectral energy distribution (SED). 
Stronger constraints on the dust temperature require additional photometry at higher frequency.
Therefore, we assume the $T_{d}$ = 47 K and $\beta$ = 1.6 for related measurements. With this assumption, the luminosities derived from the grey body model are $L_{\rm FIR}$ = (3.4$\pm$0.2) $\times$10$^{13}$ $\mu^{-1}$ $L_{\odot}$ and $L_{\rm TIR}$ = (4.8$\pm$0.2) $\times$10$^{13}$ $\mu^{-1}$ $L_{\odot}$. The derived FIR luminosity is the highest for all known high redshift quasars (observed value before correcting for lensing magnification). 

We estimate the SFR from the $L_{\rm TIR}$ based on the local scaling relation from \cite{murphy11} and assuming the $L_{\rm TIR}$ is dominated by star formation. We obtain an SFR of 7080 $\mu^{-1}$ \msunyr. Any contribution from quasar to the IR luminosity will result in an overestimation of the SFR.
The dust mass based on the assumption of dust emission model is $M_{d}$ = 2.2$\pm$0.1$\times$10$^{9}$ $\mu^{-1}$ \msun, where the uncertainty is only from fitting. 
Some recent works suggest that the relatively low metallicity galaxy template Haro 11 could provide a better fit to  the IR SED of high redshift quasar\citep{lyu16,derossi18}. As a comparison, we also fit the observed data with the Haro 11 template and a AGN template from \cite{lyu16}, which results in a 1.6 times higher TIR luminosity, consistent with the systematic difference between the optically thin grey body model and Haro 11 template \citep{derossi18}. 

The VLA 3 GHz detection represents the rest-frame 22.6 GHz radio emission from quasar. We calculate the radio-loudness by adopting the definition of $R_{\rm 2500}= f_{\nu,\rm 5GHz} / f_{\nu, \mathrm 2500 \AA}$, where the $f_{\nu, \mathrm 2500 \AA}$ is the optical flux density at rest frame 2500 \AA, measured from the NIR spectrum of J0439+1634. The rest frame 5GHz flux density $f_{\nu,\rm 5GHz}$ is estimated from the VLA detection assuming a power-law radio continuum ($f_{\nu} \propto \nu^{\alpha}$). The radio-loudness is constrained to be $R$ = 0.05 with a flat continuum $\alpha = -0.06$, and $R$ = 0.17 with a steep slope of $\alpha = -0.9$, which indicates that J0439+1634 is a radio quiet quasar. 
Based on the current lensing models, if the radio emission is dominated by the quasar, the intrinsic radio emission $L_{\rm \nu, 1.4GHz}$ will be $10^{-3.9}$--$10^{-2.9}$ \lsun\ Hz$^{-1}$ assuming a slope of $\alpha = -0.06 - -0.9$. On the other hand,  based on the radio--FIR relation \citep{yun01, wang07}, the radio emission from host could be expected to be $L_{\rm \nu, 1.4GHz}$ $\gtrsim 10^{-3}$ \lsun\ Hz$^{-1}$ when assuming the $L_{\rm FIR}$ is dominated by star formation and considering the scatter of the radio--FIR relation. This result suggests that the host galaxy contributes a significant fraction of radio emission, which could be higher than or comparable to that from quasar. Currently, only one frequency detection of radio continuum and the uncertain lensing configuration limit the more detailed investigation of radio emission properties. 

\subsection{\cii\ Emission Line}
The \cii\ 158 $\mu$m emission line is detected from NOEMA observations (Figure 2) with a line peak flux density of $\sim$ 36 mJy, the brightest \cii\ line detected from any quasar at $z \gtrsim 6$ \citep[e.g.,][]{decarli18}.
We extract the spectrum from the peak pixel in the data cube and fit the emission line with a Gaussian profile. The \cii\ line tentatively shows wings that can not be matched by a single Gaussian fit (see Figure 2). 
Due to the relatively low signal-to-noise ratio (S/N) at the wings, we adopt a single Gaussian profile as the best fit. Future high S/N observations are needed to confirm and to investigate the nature of the extended wing of \cii\ line.
The redshift measured from \cii\ line is $z_\mathrm{[CII]}$= 6.5188$\pm$0.0002, which is slightly higher than the \mgii\ -based redshift ($z=6.511\pm0.003$), indicating a 311 \kms\ blueshift of the \mgii\ line. This blue shift is similar to shifts between \mgii\ and \cii\ found in other $z>6$ quasars \citep[e.g., several hundred \kms;][]{mazzucchelli17,decarli18}.
We obtain a line flux of $F_\mathrm{[CII]}$ = 11.7$\pm$0.8 Jy\,\kms, and a FWHM$_\mathrm{[CII]}$ = 328.1$\pm$18.0 \kms, corresponding to a line luminosity of $L_\mathrm{[CII]}$ = (1.2$\pm$0.1) $\times$10$^{10}$ $\mu^{-1}$ $L_{\odot}$ ($L^\prime_\mathrm{[CII]}$ = (5.7$\pm$0.4)$\times$10$^{10}$ $\mu^{-1}$ \kkmspc). 

The detection of the \cii\ line allows us to derive a $L_\mathrm{[CII]}$ -based SFR. We use the SFR--$L_\mathrm{[CII]}$ relations for high-redshift ($z>0.5$) galaxies from \citet{delooze14}: 
\begin{equation}
\mathrm{SFR}_\mathrm{[CII]}/\mathrm{M}_\sun\,\mathrm{yr}^{-1} = 3.0\times10^{-9} (L_\mathrm{[CII]}/\mathrm{L}_\sun)^{1.18},
\end{equation}
which has a systematic uncertainty of a factor of $\sim$2.5. Therefore, we obtain an observed SFR$_\mathrm{[CII]}$ of $\sim$ 1000 -- 6000 $\mu^{-1.18}$ \msunyr, slightly lower than the SFR$_\mathrm{TIR}$.
The \cii\ -to-FIR luminosity ratio $L_\mathrm{[CII]}$/$L_{\rm FIR}$ is (3.1--4.1) $\times 10^{-4}$, within the range of the $L_\mathrm{[CII]}$/$L_{\rm FIR}$ distribution of previously known $z \gtrsim 6$ quasars \citep{venemans17b,decarli18}. 

The \cii\ line also enables us to derive the mass of singly ionized carbon $M_{C^+}$. Following \cite{venemans17c}, we calculate the $M_{C^+}$ using
\begin{eqnarray}
M_{\mathrm{C}^+}/M_\odot &=& C m_\mathrm{C} \frac{8\pi k \nu_0^2}{h c^3 A} Q(T_\mathrm{ex}) \frac{1}{4} e^{91.2/T_\mathrm{ex}} L^\prime_\mathrm{[CII]} = \nonumber \\
&=&  2.92\times10^{-4} Q(T_\mathrm{ex}) \frac{1}{4} \mathrm{e}^{91.2/T_\mathrm{ex}} 
L^\prime_\mathrm{[CII]},
\end{eqnarray}
where $C$ is the conversion between pc$^{2}$ and cm$^{2}$, $m_{C}$ is the mass of a carbon atom, $A$=2.29 $\times 10^{-6}$ s$^{-1}$ is the Einstein coefficient \citep{ns81}, $Q(T_\mathrm{ex})=2+4\mathrm{e}^{-91.2/T_\mathrm{ex}}$ is the \ion{C}{2} partition function, $T_\mathrm{ex}$ is the excitation temperature and $L^\prime_\mathrm{[CII]}$ is the line luminosity in \kkmspc. We also assume the $T_\mathrm{ex}\gtrsim100$\,K \citep[e.g.,][]{meijerink07, venemans17b}. The derived $M_{\mathrm{C}^+}$ is (3.7 $\pm$ 0.3) $\times 10 ^{7} \mu^{-1}$ with $T_\mathrm{ex}=100$\,K and 3.1 $\times 10 ^{7} \mu^{-1}$ if $T_\mathrm{ex}=200$\,K.

Although the NOEMA observations do not resolve the source, the \cii\ line could provide a constraint on the dynamical mass to first order. In a dispersion-dominated system, the dynamical mass can be estimated as $M_{\rm dyn}=  3 R_{\rm [CII]}\sigma^{2} / 2G$. We adopt the velocity dispersion $\sigma$ estimated from the FWHM$_\mathrm{[CII]}$ of Gaussian fit to the emission line and assume a radius of the \cii\ line emitting region of $R_{\rm [CII]} = 2.8$ kpc, the mean value of the $z\gtrsim 6$ quasar sample from \cite{decarli18}, which results in a dynamical mass of $M_{\rm dyn} \sim 1.9 \times 10^{10}$ \msun. 
Alternatively, the \cii\ emission might originate from a thin rotating disk with an inclination angle $i$ \citep[e.g.,][]{wang13,willott15,venemans16,venemans17b}. In this case, the dynamical mass is computed as $M_{\rm dyn}$ = $R_{\rm [CII]}$(0.75 FWHM$_\mathrm{[CII]}/$sin($i$))$^{2}/G$ \citep[e.g.,][]{ho07,wang13}. 
Since we assume {\em an intrinsic radius} $R_{\rm [CII]}$ of 2.8 kpc, the host galaxy dynamical mass is estimated directly from the line width here and not affected by the value of $\mu$ (except for the potential differential lensing effect) . 
The median value of the inclination angle of $z \sim$ 6 quasars is $i = 55^{\circ}$ \citep[e.g.,][]{wang13,venemans17b}. 
Adopting this inclination angle, we obtain $M_{\rm dyn} \sim 5.9 \times 10^{10}$ \msun. 
The lensing-corrected \mgii-based black hole mass of J0439+1634 is 6.89 $\times 10^{8}$ M$_{\odot}$ (Paper I).
Thus, the $M_{\rm BH}/M_{\rm dyn}$ ratio is $\sim$ 0.011 -- 0.036, which is consistent with the measurements of $z \gtrsim 6$ quasar sample \citep{decarli18} and higher than the typical value $\sim $ 0.002 of local galaxies.

\subsection{CO Lines and Molecular Gas Mass}
\begin{figure}
\centering
\epsscale{1.2}
\plotone{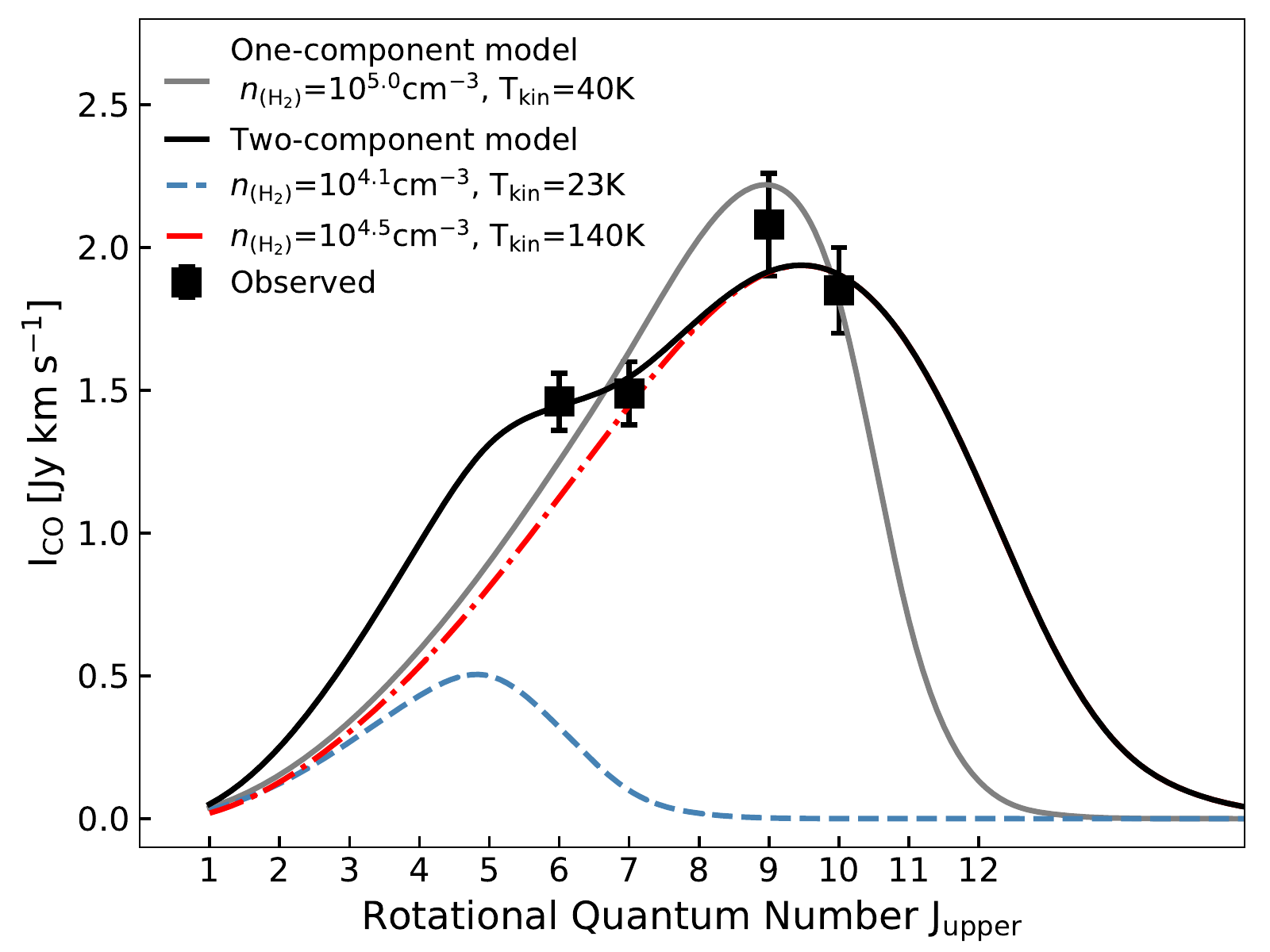} 
\caption{The COSLED of J0439+1634. Four black squares with errors are observed data from NOEMA. The grey solid line is the best-fit of a single LVG model, with $T$=40 K and $n = 10^{5} cm^{-3}$. The black solid line represents the two-component model: a ``cold'' component (blue dashed line) with $T$ = 23 K and $n = 10^{4.1} cm^{-3}$, and a ``warm'' component (red dot-dashed line) with $T$ = 140K and $n = 10^{4.5} cm^{-3}$.}
\label{fig:COSLED}
\end{figure}

With the NOEMA observations, we detect four CO emission lines: CO(6--5), CO(7--6), CO(9--8) and CO(10--9) (Figure 2).  To measure the line luminosities, we fit each emission line with a Gaussian profile. The measured line fluxes and FWHMs are summarized in Table 1. The CO(10--9) line is blended with water vapor emission H$_{2}$O $3_{\rm 1,2}-2_{\rm 2,1}$ line. We fit the blended line with two Gaussian components and fixed FWHMs. We fix the FWHM of one component to the value measured from CO(9--8) line and assume the other one to be the same FHWM to that of H$_{2}$O $3_{\rm 2,1}-3_{\rm 1,2}$ line. With multiple CO lines, we can investigate the physical condition (e.g., temperature and density) of molecular gas through the CO spectral line energy distribution (COSLED). We fit the COSLED with large velocity gradient (LVG) model using $\chi^{2}$ fitting. We use the one-dimensional non-LTE (Local Thermodynamic Equilibrium) radiative transfer code, RADEX, developed by \cite{vandertak07}. We use both a single LVG model and a two-component model for the fitting, as shown in Figure 4. We set the background temperature to be the CMB temperature at $z$ = 6.5188 of $\sim$ 20.5 K.
The best-fit of one-component model has temperature $T$ = 40 K and gas density $n = 10^{5}cm^{-3}$, while the best-fit two-component model includes a ``cold" component with $T$ = 23 K and $n = 10^{4.1} cm^{-3}$, and a ``warm" component with $T$ = 140 K and $n = 10^{4.5} cm^{-3}$. 

The current models can be used to estimate the CO(1--0) line luminosity and further constrain the molecular gas mass, although the COSLED is still uncertain due to the lack of low-$J$ CO line. 
The one-component model results in a CO(1--0) line flux of $F_{\rm CO(1-0)}$ $\sim$ 0.04 Jy\,\kms and a line luminosity of $L_{\rm CO(1-0)} \sim 2.5 \times 10^{6} L_{\odot}$ ($L^\prime_{\rm CO(1-0)} \sim 5.1 \times 10^{10}$ \kkmspc). The two-component model provides $F_{\rm CO(1-0)}$ $\sim$ 0.05 Jy\,\kms and $L_{\rm CO(1-0)} \sim 3.3 \times 10^{6} L_{\odot}$ ($L^\prime_{\rm CO(1-0)} \sim 6.7 \times 10^{10}$ \kkmspc). Therefore, the \cii\ -to- CO(1--0) luminosity ratios from one-component and two-component models are 5200 and 3900, respectively. The \cii\ -to-CO(1--0) luminosity ratio of three $z \sim 6$ quasars given by \cite{venemans17c} is 2500--4200, and the typical value measured from local and $z>2$ star forming galaxies is $\sim$ 4100 \citep{stacey91, gullberg15}. The \cii\ -to-CO(1--0) luminosity ratio derived from two-component model is more consistent with these previous measurements, although the one-component model can not be ruled out either. We adopt the two-component model to estimate the molecular gas mass $M_{\rm H_2, CO}$ using $M_{\rm H_2, CO} = \alpha$ $L^\prime_{\rm CO(1-0)}$ and assuming the luminosity-to-gas mass conversion factor of $\alpha$ =  0.8$M_{\odot}$(\kkmspc)$^{-1}$.
The molecular gas mass is estimated as 5.4 $\times$10$^{10}$ $\mu^{-1}$ \msun, consistent with the value measured from atomic gas mass (see Table 1).

\subsection{\ci\ and Atomic Carbon Mass}
The \ci\ emission line has a line luminosity of $L_\mathrm{[CI]}$ = 2.04$\pm$0.41 $\times$10$^{8} $ $\mu^{-1}$ $L_{\odot}$ ($L^\prime_\mathrm{[CI]}$ = 1.20$\pm$0.24$\times$10$^{10}$ $\mu^{-1}$ \kkmspc) and FWHM$_\mathrm{[CI]}$ = 232.04$\pm$36.59 \kms. We can estimate the atomic carbon mass based on the \ci\ line using the following relation between \ci\ brightness and the mass in neutral carbon from \cite{weiss03,weiss05} assuming the \ci\ emission is optically thin:
\begin{equation}
M_\mathrm{CI}/M_\sun= 4.566\times10^{-4} Q(T_\mathrm{ex})\frac{1}{5}e^{T_2/T_\mathrm{ex}} L^\prime_\mathrm{[CI]},
\end{equation}
where $Q(T_\mathrm{ex})=1+3e^{-T_1/T_\mathrm{ex}}+5e^{-T_2/T_\mathrm{ex}}$ is the \ion{C}{1} partition function, $T_1=23.6$\,K and $T_2=62.5$\,K are the energies above the ground state. 
We assume the excitation temperature to $T_\mathrm{ex}=30$\,K \citep[see, e.g.][]{walter11} and obtain the atomic carbon mass of $M_{\rm CI}$ = (2.6$\pm$0.5) $\times$10$^{7}$ $\mu^{-1}$ \msun.
Assuming the atomic carbon abundance (X\ci\ = $M_{\rm CI}/(6M_{\rm H_2})$) to be X\ci\ = (8.4$\pm3.5$) $\times 10^{-5}$, following \cite{walter11} and \cite{venemans17c}, we can estimate an independent molecular gas of $M_{\rm H_2, CI}$ =  (3.9 -- 8.9) $\times$10$^{10}$ $\mu^{-1}$ \msun. This is consistent with the molecular gas mass estimated from CO lines derived in the previous section.

\subsection{The Characteristics of the ISM}

\begin{figure}
\centering
\epsscale{1.2}
\plotone{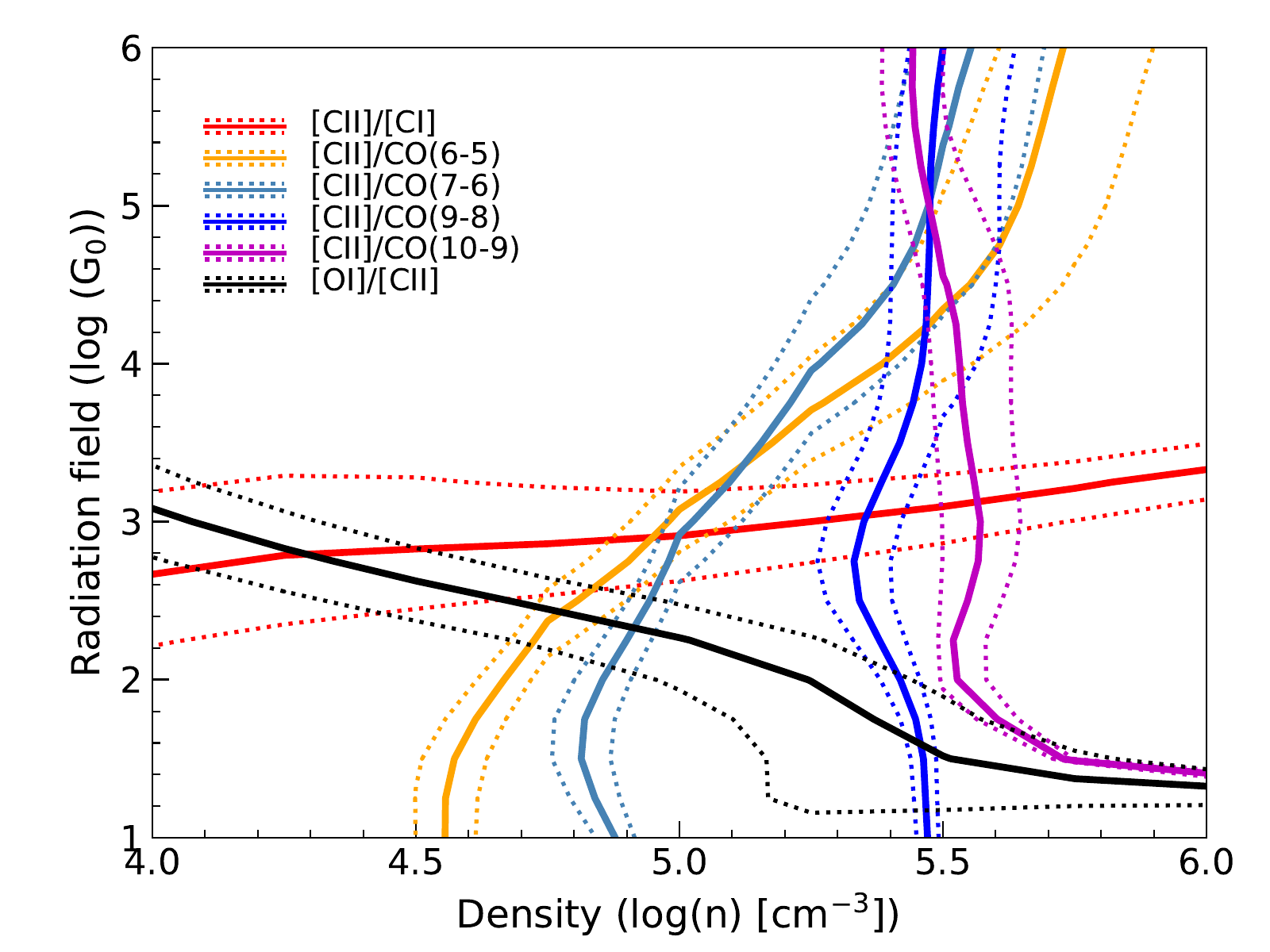} 
\plotone{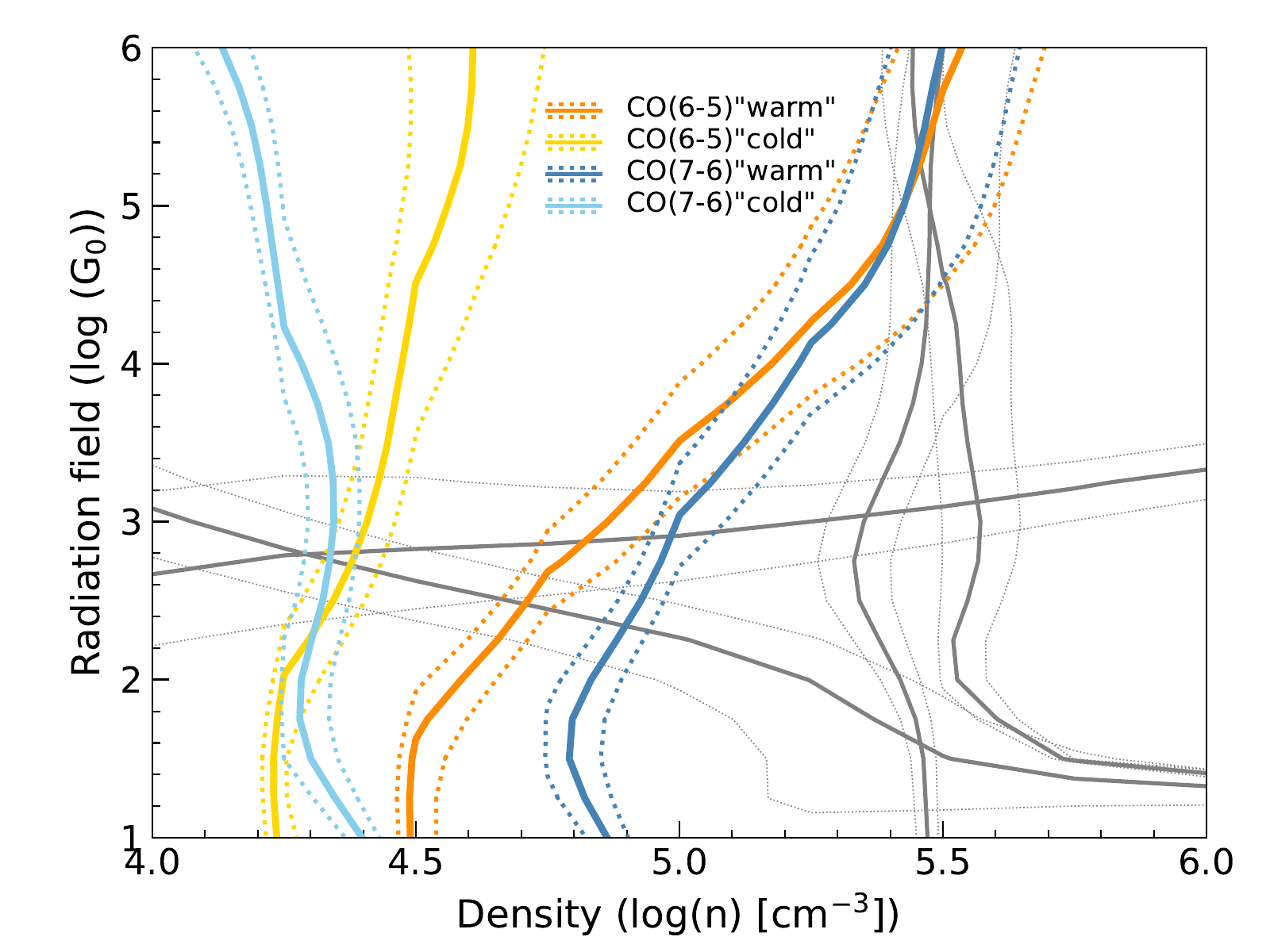}
\caption{{\bf Top:} Line luminosity ratios of $L_\mathrm{[CII]}$/$L_\mathrm{CO(6-5)}$ (=8.8, yellow), $L_\mathrm{[CII]}$/$L_\mathrm{CO(7-6)}$ (=7.4, steelblue), $L_\mathrm{[CII]}$/$L_\mathrm{CO(9-8)}$ (=4.1, blue), $L_\mathrm{[CII]}$/$L_\mathrm{CO(10-9)}$ (=4.2, purple), $L_\mathrm{[CII]}$/$L_\mathrm{[CI]}$ (=24.3, red), and $L_\mathrm{[OI]}$/$L_\mathrm{[CII]}$ (=0.3, black) of J0439+1634, compared to the predictions from the PDR model of \cite{kaufman99} in different conditions of gas density $n$ and radiation field $G$. Luminosities are all in units of $L_{\odot}$. The dotted lines are the 1$\sigma$ errors of line ratios. {\bf Bottom:} The $L_\mathrm{[CII]}$/$L_\mathrm{CO}$ of split CO(6--5) and CO(7--6) lines compared with the other four ratios (grey) shown in the top panel. We split both the CO(6--5) and CO(7--6) lines into the ``warm" and ``cold" components based on the best-fit two-component model of COLSED in Figure 4. The $L_\mathrm{[CII]}$/$L_\mathrm{CO}$ of ``warm" CO(6--5) (dark orange) and CO(7--6) (dark blue) are consistent with the radios of CO(9--8) and CO(10--9) at the region with higher $G$ and $n$, while the radios of the ``cold" CO(6--5) (yellow) and CO(7--6) (light blue) overlap with the $L_\mathrm{[CII]}$/$L_\mathrm{[CI]}$ and $L_\mathrm{[OI]}$/$L_\mathrm{[CII]}$ at a region with a lower $G$ and $n$.}
\label{fig:lineratio}
\end{figure}

The multiple FIR lines could place constraints on the physical properties of the ISM in the host galaxy through different line ratios. We compare the luminosity ratios of [\ion{C}{2}]/[\ion{C}{1}], [\ion{C}{2}]/CO and [\ion{O}{1}]/ [\ion{C}{2}] observed from J0439+1634 to the predictions by the photodissociation regions (PDR) model from PDR Toolbox \citep{kaufman99,kaufman06,pound08}. 
A one-sided illuminated slab geometry is adopted in this PDR model. In reality, assuming spherical clouds, we would detect optical thin emission from both the front and the back of the cloud, but only detect optically thick emission from the front.
Therefore, when we apply this model to observed data, we need to multiply the observed optically thick line flux by two or divide the observed optically thin emission by two to match the model. Here we divide the optically thin \cii\ line emission by a factor of two. 
We also assume the fraction of \cii\ emitted from PDR is $\sim$ 80\%, considering the dust temperature $T_{d}$ and the correlations of local luminous infrared galaxies from \cite{diazsantos17}.

The overlapped region of line ratios will constrain the gas density $n$ and the incident far-ultraviolet radiation field $G$ (in units of the Habing Field, $G_{0}$, $1.6 \times 10^{-3}$ ergs cm$^{-2}$ s$^{-1}$). 
As shown in Figure 5 (top panel), the four ratios of $L_\mathrm{[CII]}$/$L_\mathrm{CO}$ overlap in a region with high radiation field $\sim 10^{4.5}$ $G_{0}$, while the ratios of lower-$J$ CO (i.e., CO(6--5) and CO(7--6)) and ratios of higher-$J$ CO (i.e., CO(9--8) and CO(10--9)) are overlapping with $L_\mathrm{[OI]}$/$L_\mathrm{[CII]}$ and $L_\mathrm{[CII]}$/$L_\mathrm{[CI]}$ separately in regions with lower $G$ ($\sim 10^{3}$ $G_{0}$). We also try to split the CO(6--5) and CO(7--6) lines based on the best-fit two-component model of COLSED in Figure 4, as shown in Figure 5 (Bottom panel). The ``warm" components of CO(6--5) and CO(7--6) overlap with CO(9--8) and CO(10--9) at the region with high radiation field ($\sim 10^{5}$ $G_{0}$) and high density ($\sim 10^{5.5}$ $\rm cm^{-3}$), while the ``cold" components are overlapping with the ratios of $L_\mathrm{[OI]}$/$L_\mathrm{[CII]}$ and $L_\mathrm{[CII]}$/$L_\mathrm{[CI]}$ at the region having lower radiation field ($\sim 10^{3}$ $G_{0}$) and lower density ($\sim 10^{4.3}$ $\rm cm^{-3}$). This result could be explained by a two-component model of CO lines, with a ``cold" component dominated by the star formation and a ``warm" component from other heating sources (e.g. AGN).
However, we note that the simplified, one-cloud models of the ISM adopted here can be insufficient to encapsulate the full complexity of the ISM.
In addition, the magnifications of different emission lines can introduce further complications. 

\subsection{Water Vapor Emission}
Water vapor emission lines have been detected from $z > 3$ galaxies and quasars, \citep[e.g.][]{vanderwerf11, combes12, omont13, riechers13}. However, for $z > 6$ quasars, only a tentative detection of two blended water emissions has been obtained \citep{banados15}. 
From J0439+1634, two water vapor emission lines, H$_{\rm 2}$O $3_{\rm 1,2}-2_{\rm 2,1}$ and H$_{\rm 2}$O $3_{\rm 2,1}-3_{\rm 1,2}$, are detected from the NOEMA observations with high S/N (see Figure 2). 
This is the first high quality detection of multiple water vapor emission lines from quasar host galaxy at $z = 6.5$. 

A tight correlation between submm H$_{2}$O lines ($L_{\rm H_2O}$) and total infrared luminosity ($L_{\rm TIR}$) has been found in local and high redshift infrared galaxies \citep{yang13,yang16}, although the relation between submm H$_{2}$O emission and AGN activity is still unclear. 
\cite{yang16} suggested a correlation between H$_{\rm 2}$O $3_{\rm 2,1}-3_{\rm 1,2}$ line luminosity and total infrared luminosity as $L_{\rm H_2O}$/($10^{7} L_{\odot}$) = ($L_{\rm TIR}$/($10^{12} L_{\odot}$))$^{1.06\pm0.22}$, using a sample of local ultra-luminous infrared galaxies (ULIRGS) and high-redshift ($z \sim 2-4$ ) ULIRGS or hyper-luminous infrared galaxies.
The luminosity of H$_{\rm 2}$O $3_{\rm 2,1}-3_{\rm 1,2}$ detected of J0439+1634 is 7.4$\pm 1.4 \times 10^{8} \mu^{-1} L_{\odot}$. The ratio of $L_{\rm H_2O}$/$L_{\rm TIR}$ follows the observed relation within scatter, as shown in Figure 6.
 
Previous studies \citep{vanderwerf11,yang13,yang16} also show that strong-AGN-dominated sources may have smaller average ratios than those in star-forming-dominated galaxies. This could be explained by the very warm dust heated by the AGN. The higher dust temperature yields a higher IR luminosity, with only a modest increase in the rate of $\ge$ 75 $\mu$m photons which dominates the excitation of $J \le 3$ H$_{2}$O emission lines \citep{kirkpatrick15, yang16}.
The $L_{\rm H_2O(3_21-3_12)}$/$L_{\rm TIR}$ ratio of J0439+1634 is 15.4 $\times 10^{-6}$, slightly above the value based on the correlation of Hy/ULIRGS as shown in Figure 6 and higher than the mean values of both local AGN-dominated sources (6.7$\times 10^{-6}$) and local star-forming-dominated galaxies (10.8 $\times 10^{-6}$) from \cite{yang16}, which could indicate a significant contribution from star formation to the IR radiation of J0439+1634.

\begin{figure}
\centering
\epsscale{1.25}
\plotone{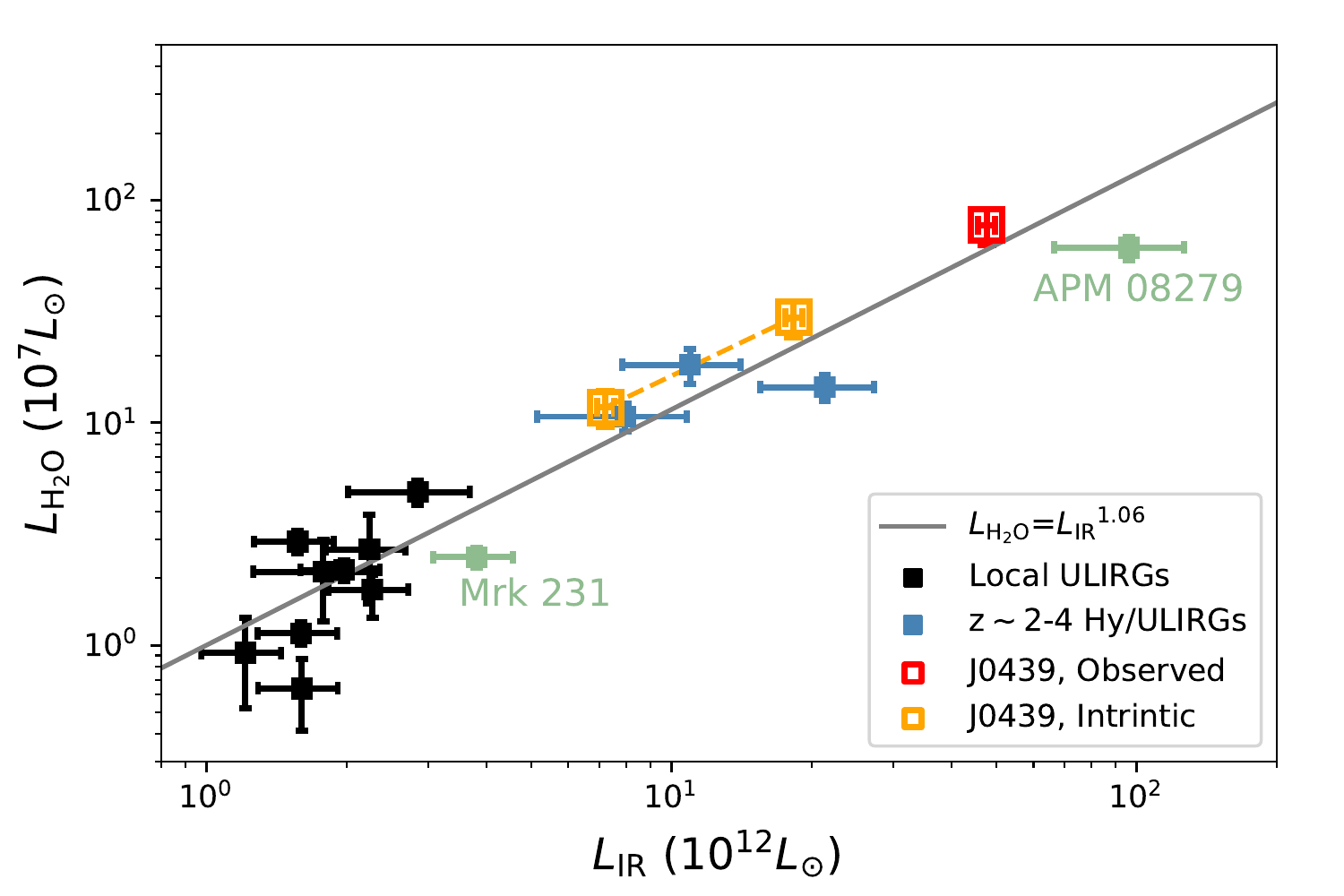} 
\caption{The luminosities of H$_{\rm 2}$O $3_{\rm 2,1}-3_{\rm 1,2}$ line and IR compared with the correlation (grey line) between $L_{\rm IR}$ and $L_{\rm H_2O 321-312}$ of local ULIRGs (black squares) and high-redshift Hy/ULIRGs (blue squares) form \cite{yang13} and \cite{yang16}. The two green squares are sources excluded from correlation fitting in \cite{yang16} due to heavy AGN contamination. We plot them here for comparison. The red open square represent the $L_{\rm IR}$ and $L_{\rm H_2O 321-312}$ before the magnification correction (observed), while the orange squares and dashed line denote the intrinsic $L_{\rm IR}$ and $L_{\rm H_2O 321-312}$ assuming the $\mu_{\rm host}$ is 2.6 -- 6.6. As shown, J0439+1634 is following this correlation.}
\label{fig:H2O-LIR}
\end{figure}

\section{Summary} 
We report FIR and radio observations of the gravitationally lensed quasar J0439+1634 at $z=6.5$. Observations from NOEMA and JCMT have detected its dust continuum emission, multiple CO, \ci, \cii, \oi, and water vapor emission lines with very high quality. Their properties make the host of J0439+1634 the brightest quasar host known at high redshift in the FIR. The VLA 3 GHz imaging detects the radio continuum of J0439+1634 at rest frame 22.6 GHz. All these observations are spatially unresolved. The main results based on these observations are summarized as following.
\begin{itemize}

\item Based on multi-frequency continuum emission detected from NOEMA and JCMT, we model J0439+1634's dust emission and estimate its FIR luminosity of $L_{\rm FIR}$ = (3.4$\pm$0.2)$\times$10$^{13}$ $\mu^{-1}$ $L_{\odot}$ with $\mu$ = 2.6--6.6, TIR luminosity of $L_{\rm TIR}$ = (4.8$\pm$0.2)$\times$10$^{13}$ $\mu^{-1}$ $L_{\odot}$, SFR$_\mathrm{TIR}$ = 7080 $\mu^{-1}$ \msunyr, and dust mass of $M_{\rm d}$ = (2.2$\pm$0.1)$\times$10$^{9}$ $\mu^{-1}$ \msun. 

\item The \cii\ line detected of J0439+1634 is the brightest \cii\ line detected from $z > 6$ source, with a peak flux density of $\sim$ 36 mJy and line luminosity of $L_\mathrm{[CII]}$ = (1.2$\pm$0.1) $\times$10$^{10}$ $\mu^{-1}$ $L_{\odot}$. The \cii\ line also provides en estimates of $L_\mathrm{[CII]}$ -based SFR, SFR$_\mathrm{[CII]}$ = 1000 -- 6000 $\mu^{-1.18}$ \msunyr and the mass of singly ionized carbon, $M_{\rm C^+}$ = (3.7$\pm$0.3) $\times 10 ^{7} \mu^{-1}$ \msun. The \cii\ line also allows us to constrain its dynamical mass to first order. 

\item With the four CO lines, CO(6--5), CO(7--6), CO(9--8), and CO(10--9), we model the COSLED of J0439+1634 with a LVG model and find that a two-component excitation model is needed to explain the observations of its molecular gas, although the constrain still has large uncertainty due to the lack of low-$J$ and very high-$J$ emission lines. Using the two-component model, we estimate the CO(1--0) line luminosity and further constrain the molecular gas mass to be $M_{\rm H_2, CO}$ = 5.4 $\times$10$^{10}$ $\mu^{-1}$ \msun.

\item The \ci\ line in J0439+1634 allows us to estimate the atomic carbon mass of $M_{\rm CI}$ =  (2.6$\pm$0.5)$\times$10$^{7}$ $\mu^{-1}$ \msun and offers an independent constraint on the molecular gas mass, $M_{\rm H_2, CI}$ = (3.9 -- 8.9) $\times$10$^{10}$ $\mu^{-1}$ \msun. We find that the molecular gas masses of J0439+1634 derived from CO(1--0) luminosity and atomic carbon mass are consistent. 

\item The line luminosity ratios between [\ion{C}{2}], [\ion{C}{1}], CO, and [\ion{O}{1}] places a first constraint on the radiation field and gas density of ISM in the host galaxy of J0439+1634. Comparing with the the predicted luminosity ratios from the PDR model of \cite{kaufman99}, to explain the result of J0439+1634, a PDR model including more than one component is required.

\item The $L_{\rm H_2O (3_{2,1}-3_{1,2})}$/$L_{\rm TIR}$ luminosity ratio of J0439+1634 follows the $L_{\rm H_2O (3_{2,1}-3_{1,2})}$--$L_{\rm TIR}$ correlation of ULIRGS and is higher than the mean value of AGN-dominated sources, probably suggesting a signification contribution from star-formation to FIR luminosity. 

\item The VLA 3 GHz observations indicate that J0439+1634 is a radio quiet quasar with radio loudness $R$ = 0.05 -- 0.17.

\end{itemize}

Future high resolution ALMA observations of J0439+1634 will fully resolve the \cii\ and dust continuum emission of the quasar host galaxy and will refine the lensing model, allowing accurate measurements of the host dynamical mass and star formation surface density. Further high resolution imaging of multiple-$J$ CO emissions could spatially resolve the excitation condition of the molecular gas. 
In addition, future observations of multiple fine structure lines (e.g., \oi, [OIII], [NII]) and molecular emissions (e.g., low- and high-$J$ CO, HCN) will provide detailed measurements of ISM properties such as excitations and metallicities in the host galaxy of J0439+1634.

\acknowledgments
We thank Jianwei Lyu, George Rieke, Chentao Yang, and Chris Carilli for important discussions.
J. Yang, X. Fan and M. Yue acknowledge support from the US NSF grant AST 15-15115 and NASA ADAP Grant NNX17AF28G.
B.P.V. and F.W. acknowledge funding through the ERC grant ``Cosmic Gas''.
C. Keeton acknowledges support from US NSF grant AST-1716585.
X.-B. Wu thanks the support by the National Key R\&D Program of China (2016YFA0400703) and the National Science Foundation of China (11533001, 11721303). 
We are grateful to the JCMT and VLA for providing DDT observations. This work is based on observations carried our under project number S18DO and W18EI with the IRAM NOEMA Interferometer. 
IRAM is supported by INSU/CNRS (France), MPG (Germany), and IGN (Spain).
The National Radio Astronomy Observatory is a facility of the National Science Foundation operated under cooperative agreement by Associated Universities, Inc.


\vspace{5mm}
\facilities{IRAM:NOEMA, JCMT, VLA}

\end{document}